\documentclass[noarxiv,a4paper,twocolumn,11pt,accepted=2026-03-18]{quantumarticle}
\pdfoutput=1
\usepackage[utf8]{inputenc}
\usepackage[english]{babel}
\usepackage[T1]{fontenc}
\usepackage{amsmath}
\usepackage{hyperref}

\usepackage{tikz}
\usepackage{lipsum}
\usepackage{comment}
\usepackage[numbers, sort&compress]{natbib}
\usepackage{graphicx}
\usepackage{dcolumn}
\usepackage{bm}
\usepackage{setspace}
\usepackage{amssymb}
\usepackage{amsthm}
\usepackage{mathrsfs}
\usepackage{bbold}
\usepackage{physics}
\usepackage{amsbsy}
\usepackage{soul, xcolor}
\setstcolor{blue}

\renewcommand{\selectlanguage}[1]{}

\newcommand{\addedtext}[1]{{#1}}
\newcommand{\fix}[1]{{#1}}
\begin{document}

\title{The Integral Decimation Method for Quantum Dynamics and Statistical Mechanics}

\author{Ryan T. Grimm}
\author{Alexander J. Staat}
\author{Joel D. Eaves}
 \email{Joel.Eaves@colorado.edu}
\affiliation{%
 Department of Chemistry, University of Colorado Boulder, Boulder, CO 80309, USA.
}%

\date{\today}

\begin{abstract}
The solutions to many problems in the mathematical, computational, and physical sciences often involve multidimensional integrals. A direct numerical evaluation of the integral incurs a computational cost that is exponential in the number of dimensions, a phenomenon called the curse of dimensionality. The problem is so substantial that one usually employs sampling methods, like Monte Carlo, to avoid integration altogether. \addedtext{Here, we derive and implement a quantum-inspired algorithm to decompose a multidimensional integrand into a product of matrix-valued
functions — a spectral tensor train — changing the computational
complexity of integration from exponential to polynomial. The algorithm constructs a spectral tensor train representation of the integrand by applying a sequence of quantum gates, where each gate corresponds to an interaction that involves increasingly more degrees of freedom in the action.} 
Because it allows for the systematic elimination of small contributions to the integral through decimation, we call the method integral decimation. The functions in the spectral basis are analytically differentiable and integrable, and in applications to the partition function, integral decimation numerically factorizes an interacting system into a product of non-interacting ones. We illustrate integral decimation by evaluating the absolute free energy and entropy of a chiral XY model as a continuous function of temperature. We also compute the nonequilibrium time-dependent reduced density matrix of a quantum chain with between two and forty levels, each coupled to colored noise. When other methods provide numerical solutions to these models, they quantitatively agree with integral decimation. When conventional methods become intractable, integral decimation can be a powerful alternative.

\end{abstract}
\maketitle

\section{Introduction}
The curse of dimensionality, where computational effort scales rapidly with the dimensionality of a problem, is a central obstacle in computational science. It arises across physics and chemistry, including in quantum many-body physics, statistical mechanics, and quantum chemistry, limiting the application of conventional, classical algorithms. The path integral acts as a versatile encoding of the behavior of physical systems, but direct solutions to high-dimensional integrals are exponentially hard to compute. Indeed, the rise of quantum computing is, in part, predicated on its ability to solve these problems. Even though quantum hardware is not yet fully developed or widely available, the study of quantum algorithms has already generated many fruitful approaches to high-dimensional problems in quantum and classical physics. \addedtext{Tensor networks facilitate these objectives by allowing classical computers to simulate quantum circuits and algorithms.}

Tensor networks compress high-dimensional tensors into contractions over low-dimensional ones and often convert the scaling of simulations from exponential to linear. In large part due to their ability to solve high-dimensional integrals, they have successfully solved a wide array of problems in computational physics, including simulating quantum circuits \cite{Seitz2023-mc}, finding the ground state of quantum many-body systems \cite{White1992-mg}, simulating open quantum systems \cite{Strathearn2018-sg}, solving stochastic \cite{Grimm2024-yg} and partial differential equations \cite{Gourianov2025-ci}, and computing the thermodynamics of lattice models \cite{Levin2007-bs}. \addedtext{The spectral tensor train (STT) decomposition, in particular, has proven to be a highly effective parameterization of multidimensional integrals \cite{Bigoni2016-kn, Grimm2024-vq, Soley2022-fw}. Once one makes the ansatz that the many-body state can be decomposed into a tensor train, one needs an accurate and efficient method of finding it.} 

Evaluating multidimensional integrals by representing them as tensor networks has proven to be accurate and efficient, even when the functions are oscillatory \cite{Nunez-Fernandez2025-wm, Khoromskij2016-em}. There are a variety of methods to find the tensor network representation of a tensor or function \citep{Ng2023-zl,Strathearn2018-sg, Verstraete2006-jc,Bloch2021-gb,Luo2023-so,Song2022-qt,Chen2022-zl,Yu2014-xh,Nunez_Fernandez2022-nj,Nunez-Fernandez2025-wm,Dolgov2020-zr,Oseledets2010-rq,Savostyanov2011-ye,Gorodetsky2018-sf,Soley2022-fw,Holtz2012-vp,Khoromskij2011-cf,Tindall2024-nd,Greene2017-lx,Grimm2024-vq,Akiyama2021-vy, Soley2022-fw, Greene2017-lx, Khoromskij2016-em}. The most general method for integration based on tensor networks is tensor cross interpolation \cite{Oseledets2011-ni, Oseledets2013-zq, Nunez-Fernandez2025-wm} (TCI), a data-driven approach that constructs an MPS by repeatedly sampling the function on the grid. TCI has been used in a wide range of applications from quantum many-body phenomena \cite{Nunez_Fernandez2022-nj}, to finance \cite{Sakurai2024-xx}, and to partial differential equations \cite{Dolgov2023-ze}. The central advantage of TCI is that it can, in principle, compute the MPS decomposition of any function without knowledge of its structure. Although TCI can achieve more rapid convergence than Monte Carlo methods \cite{Nunez_Fernandez2022-nj}, it may require a large number of function evaluations when the number of independent variables of the function is large \cite{Bigoni2016-kn}.

Eliminating small contributions to integrals through systematic recursion is a hallmark of the renormalization group, as in Kadanoff's decimation procedure \cite{Kadanoff1975-wt} and the time evolving block decimation (TEBD) method \cite{Vidal2004-kk, Tamascelli2015-rn}. Hence, we call our method integral decimation (ID). ID is a generalization and natural extension of our earlier work on both stochastic differential equations \cite{Grimm2024-yg} and open quantum systems \cite{Grimm2024-vq}.

ID turns an interacting system into a product of non-interacting ones \cite{Oseledets2010-rq, Oseledets2011-ni, Oseledets2013-zq, Nunez-Fernandez2025-wm, Nunez_Fernandez2022-nj}, just as a mean-field theory does. Such mappings are cornerstones of theoretical analysis that, for example, transform strongly interacting elementary excitations into weakly interacting quasiparticles. In rare cases, like bosonization, one can eliminate interactions altogether \cite{Senechal2003-gh}. ID seeks to transform a system of interacting particles into a non-interacting one numerically, without relying on the intuition of the theorist. Such a mapping is only practical if one can find an STT with a numerically manageable bond dimension $\chi$ of the spectral cores. 

In our earlier work on a quantum relaxation problem for the reduced density matrix in contact with a heat bath, we developed Q-ASPEN, another STT-based method \cite{Grimm2024-vq}. In that paper, the spectral cores came from optimization and importance sampling. Solving the optimization problem with stochastic gradient descent limited the scope of the method to weak and intermediate system-bath coupling strengths. In this paper, we find the spectral tensor cores through ID and remove the optimization bottleneck in our previous implementation of Q-ASPEN \cite{Grimm2024-vq}. 

But ID has applications to many problems, not just to quantum relaxation. As a simple illustration\fix{,} we use methods of ID to compress a two-dimensional Gaussian function.  Next, to demonstrate the ability of ID to compute manifestly non-Gaussian integrals, we use it to calculate the partition function of a chiral XY model.  This class of lattice models describes helical ordering in liquid crystals and helimagnets and is generally not analytically solvable \cite{D-hulst1998-su,Roy1998-au,Kousaka2009-mm}. Chiral XY models display a wide variety of winding phases, critical points, and transitions between them \cite{Yokoi1988-gw}. In the parameter ranges explored, ID is arbitrarily accurate with respect to transfer matrix solutions. Because ID allows us to compute the partition function directly, we calculate absolute free energy and entropy and compute the specific heat and internal energy through analytical differentiation of the STT.

Finally, revisiting Q-ASPEN with ID, we compute the dynamics of a non-Markovian open quantum system, a $d$-site quantum chain model. This application exhibits a complementary set of challenges to the XY-model. The weight is Gaussian but the action contains long-ranged interactions from non-Markovian system-bath correlation functions. Although the integral is Gaussian, one cannot compute it analytically, and unlike the XY model, no equivalent transfer matrix solution exists. For small numbers of levels ($d = 4$), numerically exact results are possible using the hierarchical equations of motion (HEOM) method. Here, ID agrees nearly exactly with HEOM. Circumventing the optimization bottleneck that previously limited Q-ASPEN, we present results for a $d = 40$ site quantum chain, a calculation beyond the reach of most numerically exact quantum dynamics methods. 

\section{Preliminaries}

\addedtext{In this manuscript, we focus on one-dimensional path integrals as tensor network methods for higher-dimensional systems are underdeveloped. In particular, when a network contains loops, algorithms for finding the cores---such as TCI---that rely on a one-dimensional topology break down, making the adaptation of integral decimation a key direction for future work. There are many important problems where the total space-time dimension is one,} and where exact solutions are elusive. Examples appear in classical statistical mechanics, open quantum systems \cite{Feynman1963-ax}, and stochastic differential equations \cite{Grimm2024-yg, Martin1973-xl}. Most generally, these problems involve computing path integrals of the form
\begin{equation}
    \langle \mathscr{O} (\zeta) \rangle= \int \mathcal{D}  \psi \; \mathscr{O}[\zeta, \psi] e^{i\mathcal{S}[\psi, \zeta]}
    \label{eq:functional_integral},
\end{equation}
where $\mathscr{O}$ is a local observable,  $\mathcal{S}$ is the action functional or Hamiltonian for the scalar system field $\psi$, and $\zeta$ is a parameter field not integrated out. For example,  when ${\mathscr{O}} = e^{\int d \mathsf{z} \; \zeta(\mathsf{z})\psi(\mathsf{z})}$, the result is a generating functional, central to theories of correlation and response functions. When the action corresponds to a free theory or is Gaussian,  Eq. \ref{eq:functional_integral} has exact solutions. In other cases, one turns to approximate analytical solutions or numerical simulations. 

To facilitate a numerical treatment of Eq. \ref{eq:functional_integral}, we discretize the integrand onto an $N$-point grid. Choosing a one-dimensional space-time grid $(\mathsf{z}_1,\cdots, \mathsf{z}_N)$, the discretized system field variables $\psi_n  \equiv \psi(\mathsf{z}_n)$, with $\bm \psi = (\psi_1,\psi_2,\cdots,\psi_N)$, and parameter field variables $\zeta_n = \zeta(\mathsf{z}_n)$ become fluctuations on the lattice of points. Eq. \ref{eq:functional_integral} becomes 
\begin{equation}
    \langle \mathscr{O}(\bm \zeta) \rangle  \approx \int \left[\prod_{n = 1}^N  d \psi_n \mathscr{O}_n( \psi_n, \zeta_n) \right] e^{i\mathcal{S}(\bm \psi, \bm \zeta)}.
    \label{eq:discrete_fintegral}
\end{equation}

With the domain of integration represented on a multidimensional quadrature grid, the integral is a sum over the indices of a tensor with dimension $q^N$, where $q$ is the number of quadrature points in each dimension. When $N$ is larger than about 10, it is not possible to store the tensor in memory and perform direct quadrature. Additionally, one often uses Eq. \ref{eq:functional_integral} to compute quantities like the generating function or partition function that must remain differentiable with respect to the parameter fields or temperature. Finding a numerical treatment of \addedtext{Eq. \ref{eq:discrete_fintegral}} that is accurate in high dimensions and that also preserves analytical differentiation inside
the integral remains a steep challenge. In this work we address this challenge by evaluating Eq. \ref{eq:discrete_fintegral} using spectral tensor trains (STTs) \cite{Bigoni2016-kn, Gorodetsky2019-ij}, which are a continuous analog to matrix product states (MPS).

Like TCI, our approach converts the $N$-dimensional integral into $N$ one-dimensional integrals. But unlike TCI, our approach exploits the physics of interactions and mathematical properties of the integrand in Eq. \ref{eq:functional_integral}. Specifically, the action is a sum of body-ordered functions, single-particle, pairwise, three-body, and so on. As we show, the product property $e^{a+b} = e^ae^b$ of the integral weight $e^{iS}$ invites a mapping of the integral to a quantum circuit, where each gate of the circuit corresponds to a term in the body-ordered action. The gates act sequentially, allowing one to systematically eliminate small contributions to the integral through singular value decomposition (SVD). This method avoids the curse of dimensionality because it compresses each of the low-dimensional quantum gates in sequence and never stores the entire tensor in memory. \addedtext{Under no compression, each applied gate multiplies the current bond dimension by $\chi$. Thus, for a long-ranged system with all-to-all interactions---as in Q-ASPEN---the number of gates required grows as $\mathcal{O}(N^B)$ and the bond dimension grows as $\mathcal{O}(\chi^{N^B})$, where $B$ is the number of degrees of freedom involved in the interaction. In practice, compression greatly reduces the bond dimension, but the compressed size is variable and depends on the correlation length of the system. Although we treat finite systems, approaching criticality or increasing coupling strengths will---in general---enlarge the bond dimension.} The tensor that emerges from this procedure encodes the expansion of the multidimensional function in products of orthogonal spectral basis functions, allowing analytical differentiation of the result with respect to the parameter fields.

Conceptually, ID borrows ideas from mean-field theories. Suppose the weight  $\mathscr{W}(\bm \psi, \bm \zeta) = e^{i\mathcal{S}(\bm \psi, \bm \zeta)}$ can be decomposed into a product of matrix-valued functions, called a spectral tensor train, with spectral tensor cores ${\bm W}_n $ and $  \bm{z}_n$,
 \begin{equation}
     \mathscr{W}(\bm \psi, \bm \zeta) \approx \prod_{n=1}^N  \bm{z}_n(\zeta_n) \bm W_n(\psi_n).
     \label{eq:K_STT}
 \end{equation}
When decomposed, the $N$-dimensional integral in Eq. \ref{eq:discrete_fintegral} becomes a product
 \begin{equation}
         \left \langle \mathscr{O}(\bm \zeta) \right \rangle \approx \prod_{n = 1}^N \bm{z}_n(\zeta_n)  \bm{Z}_n(\zeta_n),
         \label{eq:obs}
 \end{equation}
of $N$ one-dimensional integrals, where $\bm Z_n(\zeta)\equiv\int d\psi \; \mathscr{O}_n(\zeta, \psi)  \bm W_n(\psi) $. 

\section{Integral Decimation}

\subsection{Mapping the Integral Weight to a Quantum Circuit}

\begin{figure*}
    \centering
    \includegraphics[width=1.0\linewidth]{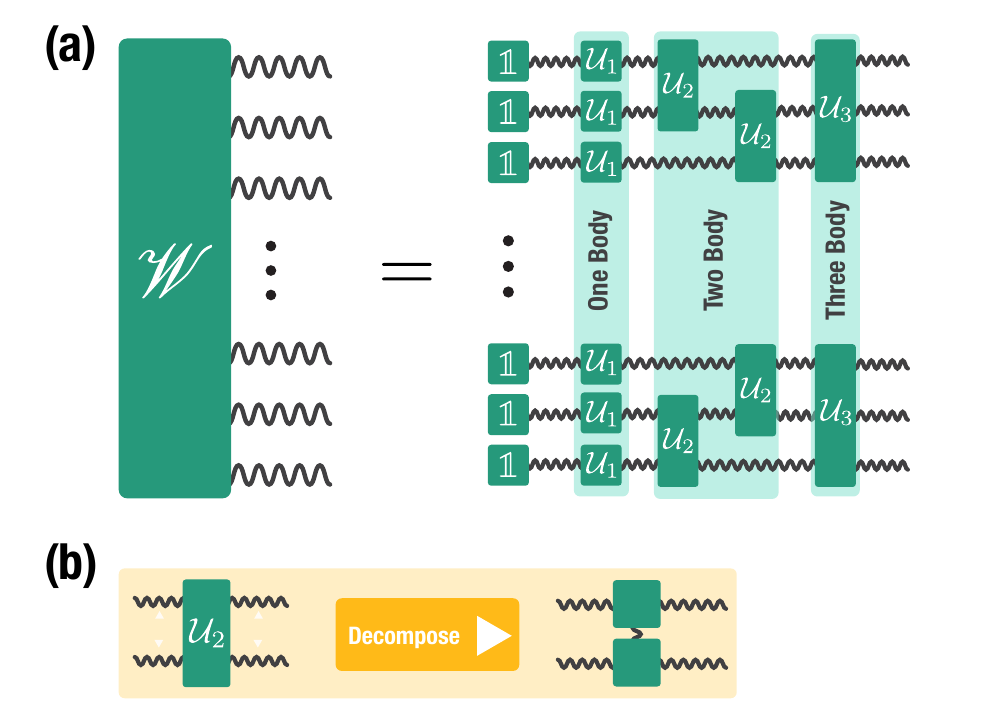}
    \caption{A diagram for the conversion of the weight $\mathscr{W}$ into a quantum circuit. (a) View the weight function as a quantum state $\ket{\mathscr{W}}$ such that $\mathcal{W}(x_1,\cdots) = \braket{\mathbf{x}}{\mathscr{W}}$. Each horizontal wavy line depicts one of the $x_n$ variables. (b) The body-ordered expansion of the action $\mathcal{S}$ corresponds to a sequence of quantum gates, where the body order of each gate corresponds to the body order of the potential. The propagation occurs along an auxiliary time-dimension (left to right), producing a tensor representation of the weight from an initially unentangled state.}
    \label{fig:decompose}
\end{figure*}

To compress the integral weight as an STT, we treat it as a many-body wavefunction. Start by defining the position-like ket $\ket{\bm{\mathsf{x}}} =  \ket{x_1, x_2, \cdots}$ where each $x_n$ corresponds to  either a system or parameter field $x_n \in \{\psi_n, \zeta_n \}$ at space-time point $\mathsf{z}_n$. The weight is the quantum amplitude 
\begin{equation}
    \mathscr{W}(\bm{\mathsf{x}}) = \braket{\bm{\mathsf{x}}}{\mathscr{W}} .
    \label{eq:weight_ket}
\end{equation}
Suppose the state $\ket{\mathscr{W}}$ is related to another state $\ket{\Psi}$ by the transformation
\begin{equation}
    \ket{\mathscr{W}} = {\mathscr  U}\ket{\Psi},
    \label{eq:qmc}
\end{equation}
because the action decomposes into a body-ordered expansion,
\begin{align}
    \mathscr{W}(\bm{\mathsf{x}}) &= e^{-i\sum_{k} V_1(x_k) -i\sum_{k,l} V_2(x_k,x_l) + \dots },  
    \label{eq:decompose}
\end{align}  
where $V_1, V_2, \dots$ are the one-body, two-body and higher interaction potentials \cite{Grimmett1973-mb, Barenco1995-yb}, it is prudent to decompose ${\mathscr U}$ into a product
\begin{equation}
   {\mathscr  U} =\prod_{\bar{n} = 1}^{\bar{N}} {\mathcal{U}}^{\bar{n}} ,
    \label{eq:evo}
\end{equation}
where each ${\mathcal{U}}^{\bar{n}}$ acts at a step $\bar{n}$.

Now insert the product form of ${\mathscr U}$ in Eq. \ref{eq:evo} into Eq. \ref{eq:qmc}, left multiply by $\bra{ \bm{ \mathsf{x} } }$, and insert position space resolutions of the identity $\mathbb{1} = \int d\bm{\mathsf{x}} \ket{\bm{\mathsf{x}}}\bra{\bm{\mathsf{x}}}$ at each time step to get
\begin{equation}
    \braket{\bm{\mathsf{x}}}{\mathscr{W}} = \int \left[\prod_{{\bar{n}}=1}^{\bar{N}} d\bm{\mathsf{x}}^{\bar{n}} \bra{\bm{\mathsf{x}}^{\bar{n} + 1}} {\mathcal{U}}^{\bar{n}} \ket{\bm{\mathsf{x}}^{\bar{n}}}  \right] \braket{\bm{\mathsf{x}}^1}{\Psi},
    \label{eq:the_point}
\end{equation} where $\bm {\mathsf{x}} ^{\bar{N}+1} = \bm {\mathsf{x}}.$
Equating Eq. \ref{eq:the_point} to Eq. \ref{eq:decompose} reveals that the initial state is the wavefunction $\braket{\bm{\mathsf{x}}}{\Psi} = 1$, which is a trivially unentangled state.  For shorthand, we define the state $\ket{\bm{1}}$ such that $\braket{\bm{\mathsf{x}}}{\bm{1}} = 1$. The state has a similarly trivial decomposition, $\ket{\bm 1} = \ket{1}\otimes\ket{1}\otimes\ket{1} \cdots $. Each ${\cal U}^{\bar{n}}$ equates to one of the terms in the sum for the action, representing a quantum gate for each term in the body-ordered expansion of the action~\cite{Nielsen2012-dp}. For example, on a one-dimensional lattice with $d$ spins interacting through a long-ranged two-body potential, there will be $\bar{N} \sim d^2$ terms in the action, each with a corresponding gate amplitude $\bra{\bm{\mathsf{x}}'} {\mathcal{U}}_2 \ket{\bm{\mathsf{x}}} = e^{-iV_2(x_k,x_{l})} \delta(\bm{\mathsf{x}} - \bm{\mathsf{x}}')$ acting on quadrature points $x_k$ and $x_l$. 

\begin{figure}
    \centering
    \includegraphics[width=1.0\linewidth]{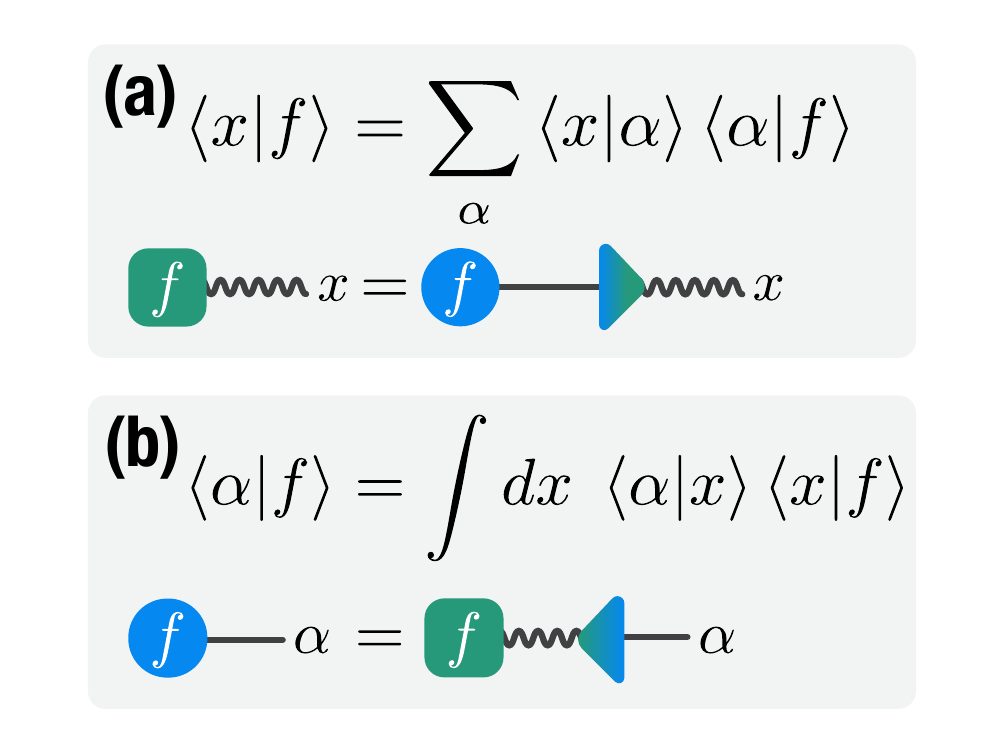}
    \caption{Converting between discrete and spectral tensor cores. Functions are green rectangles with arguments denoted by a wavy line (a), and discrete tensors are blue circles with indices denoted by a straight line (b). Triangles are transformations between the discrete and continuous representations, where contractions occur along the internal line between the shapes.  }
    \label{fig:notation}
\end{figure}

The sequence of operations mimics the time evolution of a quantum state in auxiliary time, passed through a series of mutually commuting quantum gates. The state $\ket{\bm 1}$ and the body-ordered gate functions are the key ingredients of ID. They are elements of a diagrammatic theory, shown in Figs. \ref{fig:decompose}-\ref{fig:decimation}. Each wavy line represents an argument $x_n$. The one-body functions produce a set of quantum gates ${\cal U}_1$, the two-body functions produce a set of quantum gates ${\cal U}_2$, and so on. Each element of a body-ordered function of degree $k$ has $2k$ legs in a diagrammatic representation, Fig. \ref{fig:decompose}(b). To maintain the MPS structure of $\mathscr{W}$ during propagation, we decompose gates with a body order greater than one into matrix product operators (MPO) that act on the MPS. 

\subsection{The Spectral Tensor Train Decomposition}

The spectral decomposition of the integral weight follows from a basis function expansion. Inserting a complete set of product states into Eq. \ref{eq:weight_ket},
\begin{equation}
    \bra{\bm{\mathsf{x}}}\ket{\mathscr{W}} = \sum_{\alpha_1, \alpha_2, \dots} \braket{\bm{\mathsf{x}}}{\bm \alpha} \braket{\bm \alpha}{\mathscr{W}},
    \label{eq:basis_exp}
\end{equation}
where $\ket{\bm \alpha} \equiv \ket{\alpha_1, \alpha_2, \dots}$ is a microstate and  $\braket{\bm{\mathsf{x}}}{\bm \alpha} = \phi_{\alpha_1}(x_1)\phi_{\alpha_2}(x_2)\cdots$ is a product state formed from single-particle orthonormal orbitals $\phi_\alpha(x)$ with spectral index $\alpha$. The expansion coefficients $\braket{\bm \alpha}{\mathscr{W}}$ also have an MPS decomposition $
    \braket{\bm \alpha}{\mathscr{W}} = \prod_{n} \bm A^{\alpha_n}_n$ so that the STT decomposition is a product of spectral tensor cores $\bm{\mathcal{W}}_n$,
\begin{equation}
\braket{\bm{\mathsf{x}}}{\mathscr{W}} = \prod_n \bm{\mathcal{W}}_n(x_n),
\label{eq:STT_cores}
\end{equation}
where $\bm{\mathcal{W}}_n(x_n) = \sum_{\alpha} \bm A_n^\alpha \phi_\alpha(x_n)$, and $\bm{\mathcal{W}}\in\{{\bm{z}_n, \bm{W}_n}\}$ is the spectral tensor core that corresponds to space-time point $\mathsf{z}_n$. \addedtext{Note that the product in Eq. \ref{eq:STT_cores} is ordered from right to left with increasing $n$.} With the spectral cores $\{\bm{\mathcal{W}}_n \}$ computed on the quadrature grid, the discrete tensor cores $\{ \bm A_n \}$ come from the orthogonality of the basis functions
\begin{equation}
    \bm A_n^\alpha = \int dx \;  \phi^*_\alpha(x) \bm{\mathcal{W}}_n(x). 
    \label{eq:projection}
\end{equation}
Eq. \ref{eq:STT_cores} now gives the observable $\langle \mathcal{O}(\bm \zeta) \rangle$ in Eq. \ref{eq:obs} for a set of parameter fields $\bm \zeta$. One can evaluate the observable for values of the parameters at or between any of the grid points. This is a central strength of STTs.

\subsection{Algorithm}
\begin{figure*}
    \centering
    \includegraphics[width=1.0\linewidth]{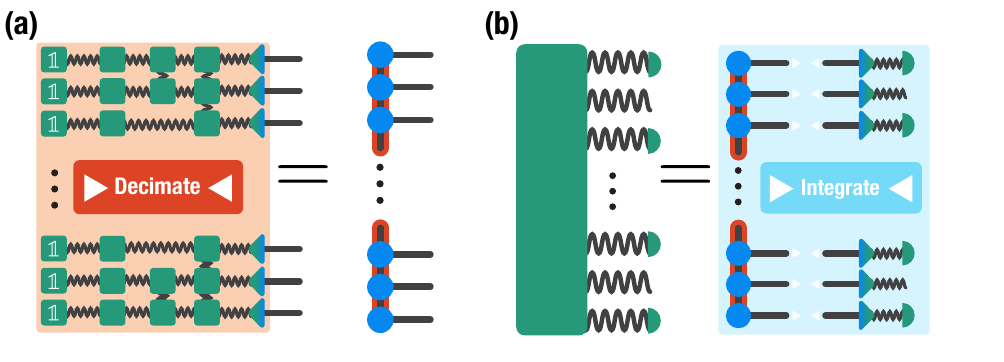}
    \caption{Constructing $\langle \mathcal{O}[\bm{\zeta}]\rangle$ in ID. (a) Contract the network layer by layer, applying the time-evolving block decimation (TEBD)  procedure sequentially. TEBD reduces the bond dimension by discarding singular values below a threshold $\epsilon_{\text{SVD}}.$  Contracting the  network results in an MPS for the coefficients of the integral weight in a spectral basis. (b) Integration (hemispherical caps) along various $x_n$ variables is straightforward. Differentiable parameter fields (uncapped wavy lines) are continuous functions in the spectral basis.}
    \label{fig:decimation}
\end{figure*}

To compute the evolution that generates $\ket{\mathscr{W}}$, we use the time-evolving block decimation procedure \cite{Vidal2004-kk, Tamascelli2015-rn} (TEBD), Fig. \ref{fig:decimation}(a). We represent the propagators as MPOs, Fig. \ref{fig:decompose}(b); apply them sequentially to the MPS; and truncate after each step. Because the propagators only act on a few degrees of freedom, they are easy to represent as MPOs, which makes TEBD efficient. The propagators have the same rank as the body order of their potential, so we can compute the MPO representation of the propagators by repeated application of singular value decomposition (SVD) \cite{Oseledets2011-ni}. \addedtext{We perform truncation using the standard algorithm~\cite{Oseledets2011-ni}, which} consists of sweeping left to right with SVD and right to left with QR decomposition.
This process brings the MPS into canonical form after the application of each gate. The last step to obtain the STT decomposition \cite{Bigoni2016-kn} is to project the state into the basis to obtain the coefficient matrices using Eq. \ref{eq:projection}. To compute the observable in Eq. \ref{eq:discrete_fintegral}, we  compute single integrals over the cores, Fig. \ref{fig:decimation}(b). 

\section{Applications}
\subsection{Gaussian} \label{sec:gauss}

\begin{figure*}
    \centering
    \includegraphics[width=1.0\linewidth]{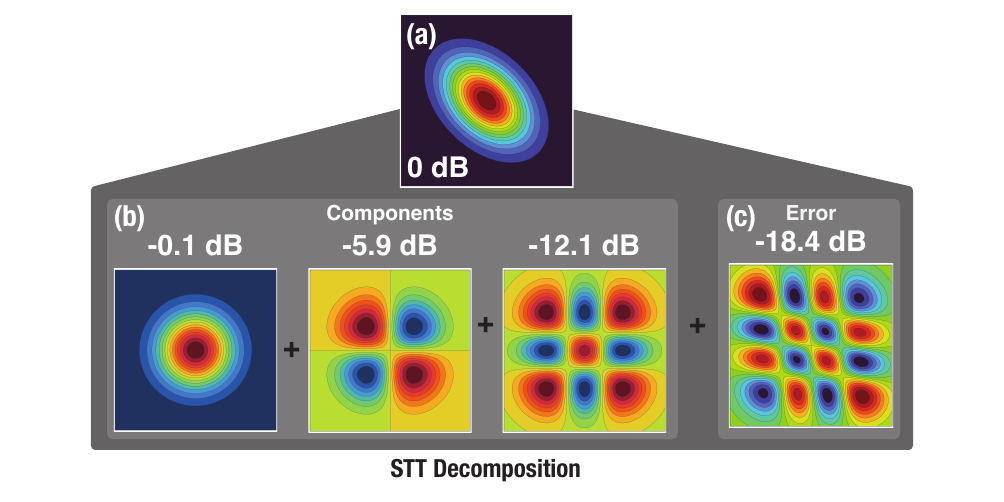}
    \caption{Illustration of the spectral tensor train (STT) decomposition of a two-dimensional, Gaussian function corresponding to the distribution function of two correlated random variables. Part (a) shows the true function where $(x,y)\in [-5,5]\times[-5,5]$. Part (b) shows the components of the STT with the three largest mean absolute magnitudes. Part (c) shows the absolute error from the true function. }
    \label{fig:gauss2d}
\end{figure*}

To illustrate the spectral decomposition in ID, we first  decompose a correlated Gaussian probability distribution into sums of separable functions. The correlated bivariate Gaussian, Fig. \ref{fig:gauss2d}(a), is
\begin{equation}
    \mathscr{W}(x,y) = \exp \left(- \frac{1}{2}\begin{bmatrix}
       x & y 
    \end{bmatrix} \bm C^{-1} \begin{bmatrix}
        x \\ y
    \end{bmatrix}\right),
\end{equation}
where $\bm C$ is a covariance matrix. An off-diagonal covariance matrix indicates that the variables (x,y) are not independent. We use ID to find a similar decomposition in terms of orthogonal polynomials. In both this and the chiral XY examples (Section \ref{sec:xy}), we employ Legendre polynomials as we have observed the best performance of the STT in rescalings of large intervals with this parameterization. In Fig. \ref{fig:gauss2d}, we decompose a distribution with a correlation matrix
\begin{equation}
    \bm C^{-1} = \begin{pmatrix}
        1/2 & 1/5 \\
        1/5 & 1/2
    \end{pmatrix}.
\end{equation} We use an SVD cutoff of $\epsilon_{\text{SVD}} = 10^{-5}$, and basis-set consisting of $b = 31$ Legendre polynomials. The mean absolute value is in decibels where $A_{dB} = 10 \log_{10}(A/A_0),$ $A$ is the mean absolute magnitude, $A_0$ is the reference magnitude, and $A_{dB}$ is the magnitude in dB. We choose the magnitude of the true function, Fig. \ref{fig:gauss2d}(a), to be the reference magnitude.

The system field variables are $\bm \psi = (x,y),$ where the system consists of two coupled sites. In this example, there are no parameters.  There are three interactions in the system: a pair of one-body, on-site interactions $V_1(x) = -i (C^{-1})_{xx}x^2 / 2$ and $V_1(y) = -i y^2 (C^{-1})_{yy} / 2$, as well as a two-body interaction $V_2(x,y) = -i x(C^{-1})_{xy}y / 2$. To represent the continuous function, first change the domain on $x$ and $y$ to have support from $-1$ to $1$. Then, expand the function in a set of orthogonal polynomials. The result is a separable decomposition 
\begin{equation}
    \mathscr{W}(x,y) = \bm W_1(x) \bm W_2(y).
\end{equation}
Because $\mathscr{W}$ is two-dimensional, we can visualize the result. The spectral cores $\bm W_1$ and $\bm W_2$ reduce to vectors, and the contraction reduces to a dot product between vector-valued functions, Fig. \ref{fig:gauss2d}(b),
\begin{equation}
    \mathscr{W}(x,y) = \sum_{j = 1}^\chi W_1^j(x) W_2^j(y),
\end{equation}
where the decomposition can be made arbitrarily accurate, Fig. \ref{fig:gauss2d}(c), by increasing the bond dimension $\chi$. \addedtext{The SVD cutoff $\epsilon_{\mathrm{SVD}}$ controls the truncation error and the bond dimension $\chi$.}
We observe a few fundamental features of ID in Fig. \ref{fig:gauss2d}(b). The weight of the basis functions decays rapidly, illustrating the compactness of the STT decomposition.  The decomposition is similar to Mehler's formula for correlated Gaussian variables. While the probability distribution is positive definite, some terms in the STT expansion are not. The STT forces an expansion of the correlated distribution function into products of separable functions, where each term is symmetric about the antidiagonal. The penalty is that not all terms in the expansion are positive definite, so that each term cannot, individually, be interpreted as a probability density.

\subsection{Chiral XY Model}  \label{sec:xy}
\begin{figure*}
    \centering
    \includegraphics[width=1.0\linewidth]{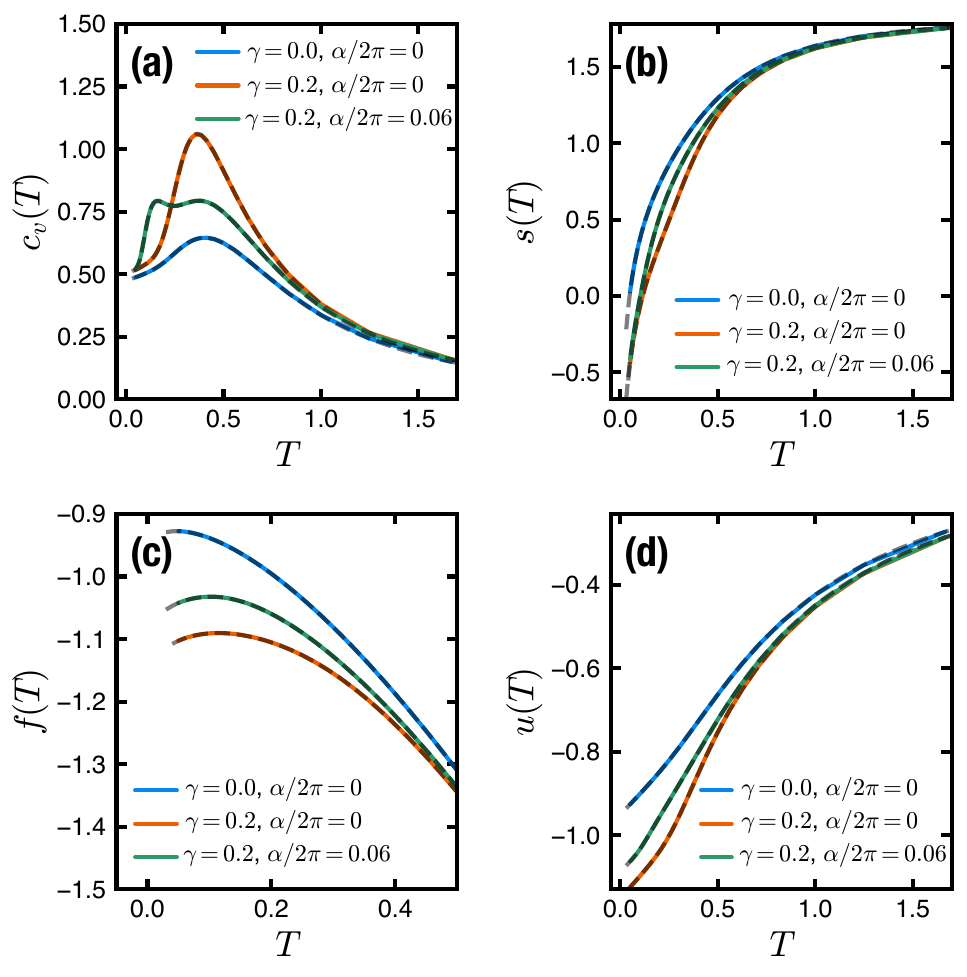}
    \caption{Thermodynamics of the classical XY model with  $d = 20$ sites as a function of temperature, field strength $\gamma$, and chirality $\alpha$. (a), (b), (c), and (d) are the specific heat, absolute entropy, free energy,  and internal energy per site, respectively. Only the specific heat and internal energy are accessible using Markov Chain Monte Carlo. Energy is in units of $J$ and temperature is in units of $J/k_B.$ Dashes correspond to benchmark solutions using a generalized transfer matrix method. The solid lines show the results obtained from the ID method: numerical integration to find the partition function as a function of temperature followed by analytical differentiation. }
    \label{fig:free_energy_XY}
\end{figure*}
To demonstrate ID on a non-Gaussian problem, we calculate thermodynamic properties of the chiral XY model, a conceptually simple one-dimensional model, by direct evaluation of the partition function. Various authors have performed numerical studies of the chiral XY model and have found rich phase and thermodynamic behavior \cite{Yokoi1988-gw, Horiguchi1998-lf, Fukui1995-op, Popov2016-ws, Horiguchi1994-wz}.  For convenience, we work in units of the coupling strength. The Hamiltonian is 
\begin{equation}
    \mathcal{H}(\bm \theta) = -J \sum_{n = 1}^{d - 1}  \cos(\theta_n - \theta_{n + 1} - \alpha) - \gamma \sum_{n = 1}^d \cos{2\theta_n},
\end{equation}
where $J > 0$ characterizes the interaction strength between adjacent spins, $\theta_n \in [-\pi,\pi)$ are the spin angles, $\gamma$ is the field strength, and $\alpha$ is the chirality parameter. A non-zero $\alpha$ introduces a bias so that adjacent spins tilt relative to one another, leading to a helical rotation along the chain. 

In the absence of a field and chirality,  there is an exact, analytical result for the partition function \cite{Mattis1984-nz, Mattis1985-un}. In the general case with chirality and an applied field, no analytical solution is known, so one must resort to numerical methods. Because the interactions are between nearest neighbors, we compute equilibrium thermodynamics using a continuous generalization of the transfer matrix method (see Appendix A) and compare to ID. 

The system field variables are $\bm \psi = (\theta_1,\dots, \theta_d)$, and the parameter field variables are the inverse temperatures that we set along the bonds between adjacent spins $\bm \zeta = (\beta_1, \dots, \beta_{d - 1}).$ Encoding the inverse temperature $\beta = 1/k_B T$ as a parameter field allows differentiation of the partition function $\cal Z$. Although the interaction potentials are two-body, it is convenient to incorporate the inverse temperature as a three-body term 
  \begin{equation}
        \begin{split}
       V_n(\theta,\beta,\theta')
= & \beta J \cos(\theta - \theta' - \alpha)
  + \beta\gamma\cos(2\theta)\\
 + &\beta\gamma\delta_{n,d-1}\cos(2\theta'),
\end{split}
  \end{equation} where the last term appears only for $n = d - 1$ in open boundary conditions. The partition function is 
\begin{equation}
    \mathcal{Z}(\bm \beta) = \int d \bm \theta \;e^{\sum_{n = 1}^{d-1} V_n(\theta_n, \beta_n, \theta_{n +1})}.
\end{equation}

Applying the STT decomposition of the weight yields
\begin{equation}
    \mathscr{W}(\bm \theta, \bm \beta) = \left[ \prod_{n = 1}^{d - 1} \bm{z}_n( \beta_n) \bm W_n(\theta_n)  \right] \bm W_d(\theta_d)
\end{equation}
so that the partition function becomes 
\begin{equation}
    \mathcal{Z}(\bm \beta) = \left[ \prod_{n = 1}^{d - 1}  \bm{z}_n( \beta_n) \bm{Z}_n \right] \bm Z_d,
\end{equation}
where $\bm Z_n = \int d\theta \; \bm W_n(\theta).$

Unlike Markov Chain Monte Carlo, ID computes the partition function, where it is possible to compute the absolute free energies and entropies. The STT is an expansion of the partition function in terms of continuous basis functions and is therefore analytically differentiable in $\beta_n$. The free energy is related to the partition function by $\beta F = -\ln(\mathcal{Z})$, and the entropy by $S/k_B = \beta(U - F).$ To compute the internal energy $U = \langle E \rangle$, we differentiate the partition function with respect to each $\beta_n$ and then take the limit that $\beta_n \rightarrow \beta$ for all $n$,
\begin{equation}
    U = -\lim_{\beta_1 \to \beta} \cdots \lim_{\beta_{d-1} \to \beta}\sum_n \frac{\partial \ln(\mathcal{Z})}{\partial \beta_n} .
\end{equation}
Intensive quantities of the internal energy $u \equiv U/d$, entropy $s \equiv S/d$, specific heat $c_v \equiv C_v/d$, and free energy $f = F/d$ for a $d = 20$ site model appear in Fig. \ref{fig:free_energy_XY}. Fig. \ref{fig:free_energy_XY} shows three cases: the exactly solvable XY model without an external field or chirality ($\gamma = \alpha = 0$), the XY model with an external field ($\gamma = 0.2$ and $\alpha = 0$), and the chiral XY model ($\gamma = 0.2$ and $\alpha / 2\pi = 0.06$). We compute the transfer matrix solution on a grid with a spacing of $\delta \theta = 2 \pi / 300.$
For ID, we use  $b = 51$ Legendre basis polynomials and an SVD accuracy of $\epsilon_{\text{SVD}} = 10^{-12}.$ The results of ID are indistinguishable from numerically exact transfer matrix solutions. 

The thermodynamics of the chiral XY model are unusual \cite{Fukui1995-op}. First, consider the temperature-dependent heat capacity in Fig. \ref{fig:free_energy_XY}(a), which shows that the specific heat tends to $k_B/2$ as $T\rightarrow0$. At low temperatures, fluctuations in $\theta$ are small enough to be Gaussian. At the lowest temperatures, a saddle point treatment of the partition function gives the result for independent classical harmonic oscillators, each of which contributes \(k_B/2\) to the heat capacity. 

Because the boundary conditions are open and not periodic, an applied field introduces one more degree of freedom. For $\gamma \neq 0$, the Hamiltonian depends on the absolute values of the angles. When $\gamma = 0$, there are $d-1$ angle differences and therefore $d-1$ degrees of freedom. But when $\gamma \neq 0$, there are $d$ angles and therefore $d$ degrees of freedom. This extra degree of freedom produces a gap in the specific heat as $T \rightarrow 0$ that would vanish as $1/d$ in the thermodynamic limit. The gap is small but visible in the specific heat near $T = 0$ even for $d = 20$, Fig. \ref{fig:free_energy_XY}(a). 

Collective excitations of massless spin defects control the heat capacity at low temperature, so that $C_v \sim T$ near $T = 0$. At high temperatures, the bounded energy spectrum causes the Boltzmann distribution to flatten to 1 and the specific heat vanishes as $1/T^2$. There must, therefore, be at least one peak at finite temperature. Peaks in the heat capacity, akin to Schottky anomalies, appear for all field strengths, Fig. \ref{fig:free_energy_XY}(a). When $\gamma$ is large enough that the external field competes with the chirality parameter $\alpha$, thermal fluctuations for angles aligned with $\alpha$ and with $\gamma$ can occur, giving rise to two peaks in $C_v$ at finite temperatures.

The entropy is logarithmic near $T = 0$ and can become negative. These results are also observable both in Fig. \ref{fig:free_energy_XY}(b) and in the work of Fukui and Horiguchi \cite{Fukui1995-op}. The Sackur-Tetrode equation for the absolute entropy of the ideal gas and the classical Heisenberg model \cite{Joyce1967-qd} both exhibit a logarithmic entropy that goes negative at low temperature. The entropy turns negative at low temperatures because the classical chiral XY model lacks the quantum fluctuations that would restore the third law of thermodynamics. 

\subsection{Non-Markovian Quantum Dynamics}  \label{sec:qaspen}

\begin{figure*}
    \centering
    \includegraphics[width=1.0\linewidth]{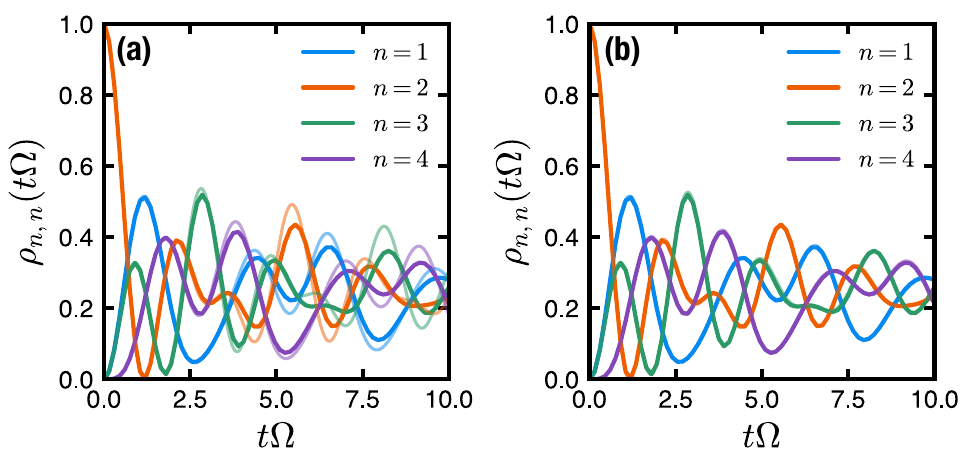}
    \caption{Population dynamics from ID compared to perturbative and numerically exact quantum dynamics methods for the quantum chain model with $d = 4$ sites described in Eqs. \ref{eq:chain_H}-\ref{eq:chain_V}, where $n$ is the site index and $\rho$ is the reduced density matrix. Part (a) shows a comparison of Q-ASPEN implemented with ID (opaque) to non-secular, second-order, time-convolutionless Redfield theory (semi-transparent). Part (b) shows a comparison of Q-ASPEN/ID (opaque) to HEOM (semi-transparent).}
    \label{fig:HEOM}
\end{figure*}

Next, we use ID to compute the dynamics of a large multi-level open-quantum system. In our recent papers on the ASPEN  \cite{Grimm2024-yg} and Q-ASPEN methods \cite{Grimm2024-vq}, we found a path integral formalism for solving stochastic differential equations with multiplicative noise,  enabling us to also phrase this problem in terms of a functional integral. In units where $\hbar = 1$, the density matrix of an open quantum system obeys a stochastic differential equation 
\begin{equation}
    i\frac{d\bm{\rho}(t)}{dt} = \bm{{\cal L}}(t) \bm{\rho}(t),
    \label{eq:liouville}
\end{equation}
where $\bm{ \mathcal{L}}(t)$ is the total, stochastic Liouvillian. This formalism can treat systems with intrinsic and extrinsic noise. However, to make contact with existing benchmarks \cite{Makri2020-wq}, we limit the structure to a Liouvillian that encodes the dynamics of a quantum subsystem coupled to a thermal bath at an inverse temperature $\beta$. In this case, the Liouvillian is $\bm{{\cal L}}(t) = \bm L_0 + \xi(t) \bm L^-_1 + \nu(t) \bm L^+_1$. $\bm L_0 \leftrightarrow [\bm H, \cdot]$ generates the dynamics of the system under a Hamiltonian $\bm H$ in the absence of fluctuations. $\bm L^-_1  \leftrightarrow [\bm V, \cdot]$ and $\bm L_1^+  \leftrightarrow \{\bm V, \cdot \}$ couple the system, via an operator $\bm V$, to a pair of Gaussian noise fields $\xi(t)$  and $\nu(t)$. The real-valued spectral density $J(\omega)$ determines the two noise sources $\xi(t)$ and $\nu(t)$, which, for thermal fluctuations, are related by a fluctuation-dissipation relation that ensures the reduced density matrix elements relax to Boltzmann equilibrium. 

The relaxation operator $\bm \Phi_M$ propagates the reduced density matrix $\langle \bm \rho(t_M) \rangle = \bm \Phi_M \bm \rho(0)$ on \fix{a} discrete time grid $t_n \equiv \tau n$, where $\tau$ is the time step. An $M$-dimensional partition sum over the eigenfrequencies $\omega^-$ and $\omega^+$ of the operators $\bm L_1^-$ and $\bm L_1^+$ specifies the relaxation operator 
\begin{equation}
    \bm \Phi_M = \sum_{\bm \omega} \left [ \prod_{n = 1}^M \bm{\mathcal{G}}_0(\omega^-_n, \omega_n^+) \right] \mathscr{W}(\bm \omega),
    \label{eq:QASPEN_integral}
\end{equation}
where $\bm{\mathcal{G}}_0$ is the free-propagator, and the path over the eigenfrequencies is  $\bm \omega = (\omega_M^-,  \omega^+_M, \dots, \omega_1^-, \omega_1^+)$. The influence weight $\mathscr{W}$ is a complex Gaussian,
\begin{equation}
     {\mathscr{W}}(\bm \omega) \equiv \exp\left(-\frac{1}{2} \bm \omega^{\intercal}
    \bm G
    \bm \omega
    \right), 
\end{equation}
that weights each path, where $\bm G$ is the correlation matrix for the noise fields $\xi$ and $\nu$. In this example, there is no parameter field, and the system field variables consist of the eigenfrequencies $\bm \psi = \bm \omega.$

Unlike the XY model, the integral in Eq. \ref{eq:QASPEN_integral} is Gaussian, but, because the integrand has an intricate dependence on the fields, we cannot convert it to a separable problem. Long-range interactions in $\bm G$ that couple the evolution of the system at different points in time make a transfer matrix solution prohibitive for $M \gg 1$. Indeed, one can truncate the interaction and form a long-ranged transfer matrix to obtain the quasi-adiabatic path-integral method (QUAPI) \cite{Makri1995-gm, Makri1995-ns}, which has poor scaling in both system size and interaction range. Writing the transfer matrix as a discrete tensor network, one arrives at the time-evolving matrix product operator method (TEMPO) \cite{Strathearn2018-sg}, which scales well in interaction range but not in system size. Writing the transfer matrix instead as a spectral tensor network yields Q-ASPEN, which, if one can find the cores, has excellent scaling in both system size and memory length. In all of these methods, the need to work with transfer matrices with a finite interaction range leads to significant computational cost and algorithmic complexity. Using ID, we can evaluate Eq. \ref{eq:QASPEN_integral} directly with little more complexity than evaluating a classical partition function. Decomposing $\mathscr{W}$ as an STT,
 \begin{equation}
     \mathscr{W}(\bm \omega) \approx \prod_{n = 1}^M \bm W^-_n(\omega^-_n) \bm W^+_n(\omega^+_n) ,
     \label{eq:QASPEN_STT}
\end{equation}
and inserting it into Eq. \ref{eq:QASPEN_integral} yields 
\begin{equation}
    \bm \Phi_M  = \prod_{n = 1}^M \bm Z_n,
    \label{eq:markov_prop}
\end{equation}
where 
\begin{equation}
    \bm Z_n \equiv \sum_{(\omega^-, \; \omega^+)} \bm{\mathcal{G}}_0(\omega^-, \; \omega^+) \otimes \bm W^-_n(\omega^-)\bm W^+_n(\omega^+)
\end{equation}
are Markovian (non-interacting) propagators. Because the Markovian propagators carry the free-propagator along with them, storing the Markovian propagators as a dense matrix is usually costly, so we represent them as two-site matrix product operators. We then efficiently contract them using zip-up \cite{Stoudenmire2010-ha} to obtain the relaxation operator.

\begin{figure}[!htb]
    \centering
    \includegraphics[width=1.0\linewidth]{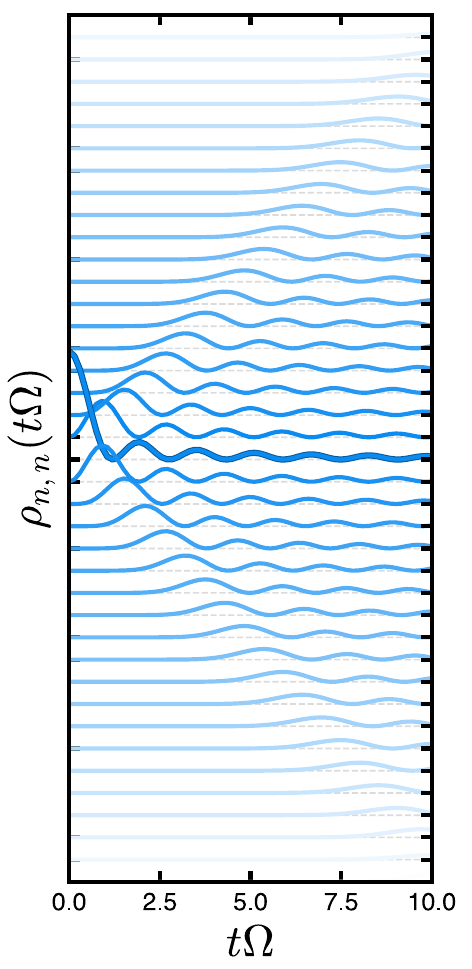}
    \caption{Population dynamics from Q-ASPEN for the quantum chain model described in the text with $d = 40$ sites. \addedtext{We use a sliding opacity to draw attention to the initialization of the wavepacket.}}
    \label{fig:many_site}
\end{figure}

To benchmark this application of ID, we simulate the dynamics of a standard benchmark system: a quantum chain model with nearest neighbor couplings \cite{Makri2020-wq}. This model occurs in many contexts, including the charge qubit Hamiltonian \cite{Pashkin2009-im}, exciton models in a single excitation subspace \cite{Holstein1959-la}, and the tight-binding model \cite{Makri2020-wq}. Working in the site basis, the system Hamiltonian is 
\begin{equation}
    H =  \Omega  \sum_{n = 1}^{d - 1}  \big(\ket{n} \bra{n + 1} + \ket{n + 1} \bra{n} \big),
    \label{eq:chain_H}
\end{equation}
where $\Omega$ is the inter-site coupling. The system-environment coupling operator is 
\begin{equation}
    V = \sum_{n = 1}^d \Delta \sigma \left(\frac{n - 1}{d - 1} - \frac{1}{2}\right)\ket{n}\bra{n},
    \label{eq:chain_V}
\end{equation}
where $\Delta \sigma$ is the radius of the chain. We couple the system to intrinsic noise with \fix{an} Ohmic spectral density $J(\omega) = \gamma \omega \exp(-\omega/\omega_c)$, where $\gamma$ is the noise strength, and $\omega_c$ is the cutoff frequency.

We benchmark the accuracy of the method at computing full memory, non-perturbative dynamics on a small system with $d = 4$ sites that is still tractable by other numerically exact methods. In Fig. \ref{fig:HEOM}(a), we compare to Redfield theory. \addedtext{As Redfield is a perturbative dynamics method that works best in the weak system-bath coupling and short bath memory limit, a choice of $\gamma = 1/2$ puts us in a dynamical regime in which Redfield will be a decent approximation at early times, as confirmed by Fig. \ref{fig:HEOM}.}
In Fig. \ref{fig:HEOM}(b), we compare to HEOM \cite{Tanimura2020-au, Tanimura1989-pa}, which is numerically exact, and find quantitative agreement, demonstrating the accuracy of ID in a challenging regime. Finally, to test the ability of ID to simulate the dynamics of a large system with no memory cutoff, we compute the dynamics of a $d = 40$-site system with identical parameters out to a time sufficiently long to see the excitation propagate to the edge of the chain, Fig. \ref{fig:many_site}. At this system size, we can observe the wavefront as it propagates through the chain.

In both Fig. \ref{fig:HEOM} and Fig. \ref{fig:many_site}, we choose system parameters $\omega_c = 1/2$, $\gamma =  1/2$, $\beta = 1$, and $\Delta \sigma = 2$, and start the excitation in the center of the chain.  In HEOM, we truncate the hierarchy at five levels and use two Matsubara frequencies. The time-step is  $dt = 0.01$ in both HEOM and Redfield. For ID, we use $b = 5$ Chebyshev basis polynomials for the STT decomposition with $\epsilon_{\text{SVD}} = 10^{-4}$ and $\tau = 0.15$. \addedtext{For the cases with exact reference calculations, we improve the convergence parameters until the desired accuracy with respect to the reference is achieved. For the forty-site system, which has no reference answer, we use the parameters from the four-site system as we do not expect that a change in system size should affect intensive parameters, such as the cutoff or basis size.}

\section{Conclusion}

The STT ansatz, Eq. \ref{eq:K_STT}, factorizes the path integral in Eq. \ref{eq:functional_integral}, transforming a high-dimensional integral into a product of one-dimensional integrals. The resulting function is analytic and differentiable in the parameter fields. ID can compute absolute quantities, like the entropy and free energy using numerical integration followed by analytical differentiation. Monte Carlo integration becomes impractical in dimensions greater than about ten. Monte Carlo also converges slowly for oscillating integrands, a phenomenon sometimes called the sign problem \cite{Nunez-Fernandez2025-wm}. Methods like ID can converge more quickly. ID also allows for a controlled separation of scales, allowing one to optimize for function interpolation, differentiation, and integration by sampling different quadrature grids at different sampling frequencies along each dimension of the integrand. We demonstrate ID in an illustrative problem to factorize a joint probability function, to compute absolute thermodynamics in an XY model with a nonpolynomial action, and to compute the relaxation dynamics in a quantum chain with non-Markovian noise as a function of the number of sites $d$, going from $d = 2$ to $d = 40$. Results from ID are indistinguishable from other numerically exact solutions when they are available.

ID exploits the decomposition of the action in terms of body-ordered functions and uses the product property of the exponential to map the weight of a path integral to a many-body wavefunction. Starting from a trivially unentangled state, ID systematically constructs a low-rank approximation to the multidimensional function by applying a sequence of quantum gates, where each gate corresponds to a body-ordered term in the action. ID avoids the memory bottleneck, or curse of dimensionality, because the dimension of each quantum gate is small enough to fit into memory. Each gate corresponds to a body-ordered term in the action with far fewer dimensions than the integrand, which allows for a straightforward compression through singular value decomposition. In one-dimensional problems, there are several methods to find tensor cores. But for tensors with more complex topologies---like projected entangled pair states \cite{Orus2014-ky}---the local, sequential, and memory-efficient approach that ID employs may be an effective way to find low-rank approximations.  

\section*{Code Availability}
All code used and data generated in the manuscript are available on \href{https://github.com/rtgrimm/integral-decimation-paper-main}{GitHub}.

\section*{Acknowledgments}
R.T.G. was supported by the National Science Foundation Graduate Research Fellowship. This material is based upon work supported by the National Science Foundation Graduate Research Fellowship Program under Grant No. (DGE 2040434). Any opinions, findings, and conclusions or recommendations expressed in this material are those of the author(s) and do not necessarily reflect the views of the National Science Foundation. This work was supported by the donors of ACS Petroleum Research Fund under New Directions Grant 68732-ND6. J.D.E. served as Principal Investigator on ACS PRF 68732-ND6 that provided support for R.T.G. A.J.S. was supported by the Air Force Office of Scientific Research under AFOSR Award No. FA9550-19-1-0083.  This work utilized the Alpine high performance computing resource at the University of Colorado Boulder. Alpine is jointly funded by the University of Colorado Boulder, the University of Colorado Anschutz, and Colorado State University and with support from NSF grants OAC-2201538 and OAC-2322260.

\section{Appendix A}
In this appendix, we derive an exact integral operator solution to the chiral XY model. The Hamiltonian is 
\begin{equation}
   \mathcal{H}(\bm \theta) = -\sum_{n = 1}^{d - 1} \cos(\theta_n - \theta_{n + 1} - \alpha) - \gamma \sum_{n = 1}^d \cos(2\theta_n). 
\end{equation}
Thus, the partition function is 
\begin{equation}
    \mathcal{Z} = \int d \bm \theta \exp(-\beta \mathcal{H}(\bm \theta)).
\end{equation}
The model is easy to solve exactly using a generalization of the transfer matrix trick. First, expand the integral to get 
\begin{equation}
    \mathcal{Z} = \int d  \bm\theta \; \left \{ \prod_{n=1}^{d-1} \mathcal{K}(\theta_n, \theta_{n+1}) \right\} \left \{ \exp(\beta \gamma \cos(2\theta_d))  \right \},
\end{equation}
where the integral kernel is $\mathcal{K}(\theta, \theta') = \exp(\beta \left[\cos(\theta - \theta' - \alpha) + \gamma  \cos(2\theta)\right])$. Rewrite the partition function by distributing the integral to obtain a generalized transfer matrix expression for the partition function 
\begin{widetext}
    \begin{equation}
    \mathcal{Z} = \int d\theta_1 \int d\theta_2 \; \mathcal{K}(\theta_1, \theta_2) \cdots \int d\theta_d \; \mathcal{K}(\theta_{d - 1}, \theta_d) \eta_d(\theta_d),
\end{equation}
\end{widetext}
where $\eta_n(\theta') = \int d\theta \; \mathcal{K}(\theta',\theta) \eta_{n+1}(\theta)$ for $n \in [1,\dots,d -1]$, and $\eta_d(\theta) = \exp(\beta \gamma \cos(2\theta))$.

\bibliographystyle{quantum}
\bibliography{references}

\end{document}